\documentclass[conference]{IEEEtran}
\ifCLASSINFOpdf
   \usepackage[pdftex]{graphicx}
\else
   \usepackage[dvips]{graphicx}
\fi
\usepackage[cmex10]{amsmath}

\usepackage{subcaption}

\usepackage{array}
\begin{document}
%
\title{On-the-fly Historical Handwritten Text Annotation}




\author{\IEEEauthorblockN{Ekta Vats and Anders Hast}
\IEEEauthorblockA{Division of Visual Information and Interaction\\
Department of Information Technology\\
Uppsala University, SE-751 05 Uppsala, Sweden\\
Email: ekta.vats@it.uu.se; anders.hast@it.uu.se}
}


%


\maketitle

\begin{abstract}

The performance of information retrieval algorithms depends upon the availability of ground truth labels annotated by experts. This is an important prerequisite, and difficulties arise when the annotated ground truth labels are incorrect or incomplete due to high levels of degradation. To address this problem, this paper presents a simple method to perform on-the-fly annotation of degraded historical handwritten text in ancient manuscripts. The proposed method aims at quick generation of ground truth and correction of inaccurate annotations such that the bounding box perfectly encapsulates the word, and contains no added noise from the background or surroundings. This method will potentially be of help to historians and researchers in generating and correcting word labels in a document dynamically. The effectiveness of the annotation method is empirically evaluated on an archival manuscript collection from well-known publicly available datasets.

\end{abstract}


%
\IEEEpeerreviewmaketitle

\section{Introduction}

Libraries and cultural organisations contain valuable manuscripts from ancient times that are to be digitised for preservation and protection from degradation over time. However, digitisation and automatic recognition of handwritten archival manuscripts is a challenging task. Unlike modern machine-printed documents that have simple layouts and common fonts, ancient handwritten documents have complex layouts and paper degradation over time. They commonly suffer from variability in writing behaviour and degradations such as ink bleed through, faded ink, wrinkles, stained paper, missing data, poor contrast, warping effects or even noise due to lighting variation during document scanning. Such issues hamper manuscript readability and pose difficulties for experts in generating ground truth annotations that are essential for information retrieval algorithms. 

Efforts have been made towards automatic generation of ground truth in the literature \cite{pletschacher2010page,heroux2007automatic,antonacopoulos2006ground}. Some popular tools include TrueViz\cite{kanungoa2001trueviz}, PerfectDoc \cite{yacoub2005perfectdoc}, PixLabeler \cite{saund2009pixlabeler}, GEDI \cite{doermann2010gedi}, Aletheia \cite{clausner2011aletheia}, WebGT \cite{biller2013webgt} and TEA \cite{valsecchi2016text}. However, most of these tools and methods are not suitable for annotating historical datasets. For example, PixLabeler and TrueViz are useful for labeling documents with regular elements or OCR text, rather than historical documents with handwritten text. Typically, these tools are hardware specific, with strict system requirements for configuration and installation. For example, TrueViz is a Java based tool for editing and visualising ground truth for OCR, and uses labels in XML format. GEDI is a highly configurable tool with XML-based metadata. Most of these methods represent ground truth with imprecise and inaccurate bounding boxes, as discussed in \cite{wei2017use}, and are suitable for a certain application. For example, PerfectDoc is commonly used for document correction instead of ground truth generation.

To address these issues, this paper presents a simple method for quickly annotating degraded historical handwritten text on-the-fly. The proposed text annotation method allows a user to view a document page, select the desired word to be labeled with simple drag-and-drop feature, and generate the corresponding bounding box annotation. In the next step, it automatically adjusts and corrects user generated bounding box annotation, such that the word perfectly fits in the bounding box and contains no added noise from the background or surroundings. 

This method can help historians and computer scientists in generating and correcting ground truth corresponding to words in a document dynamically. Also, the method can potentially benefit the research community working on document image analysis and recognition based applications where ground truth is indispensable for method evaluation. For example, in a keyword spotting scenario \cite{giotis2017survey}, if a ground truth label corresponding to a query word is missing or is inaccurate, the proposed method can be used for quick generation or correction of annotations on-the-fly.

This paper is organized as follows. Section \ref{sec:lit} discusses various document image annotation methods available in literature. Section \ref{sec:method} explains the proposed word annotation algorithm in detail. Section \ref{sec:experiments} demonstrate the efficacy of the proposed method on two well-known historical document datasets. Section \ref{sec:conclusion} concludes the paper.

\section{Document annotation methods and tools}
\label{sec:lit}

Several document image ground truth annotation methods and tools have been suggested in literature. An approach for automatic generation of ground truth for image analysis systems is proposed in \cite{heroux2007automatic}. Generated ground truth contains information related to layout structure, formatting rules and reading order. An XML-based page image representation framework PAGE (Page Analysis and Ground-truth Elements) was proposed in \cite{pletschacher2010page}. Problems related to ground truth design, representation and creation are discussed in \cite{antonacopoulos2006ground}. However, these methods are not suitable for annotating degraded historical datasets with complex layouts.

Ground truth can be automatically generated using popular tools. The earliest tool was Pink Panther \cite{yanikoglu1998pink} that allows a user to annotate document image with simple mouse clicks. TrueViz\cite{kanungoa2001trueviz} is a commonly used Java based tool for editing and visualising ground truth for OCR text, and is not suitable for handwritten historical documents annotation. PerfectDoc \cite{yacoub2005perfectdoc} is a tool used for document correction, with possible application to ground truth generation. One of the most popular document labeling tools is PixLabeler \cite{saund2009pixlabeler}. It uses a Java based user interface for annotating documents at pixel level. Like Pink Panther, TrueViz and PerfectDoc, PixLabeler works well on simple documents only and perform poorly on historical handwritten document images \cite{wei2017use}. The tool GEDI \cite{doermann2010gedi} supports multiple functionalities such as merging, splitting and ordering. It is a highly configurable document annotation tool. The advanced tool Aletheia is proposed in \cite{clausner2011aletheia} for accurate and cost effective ground truth generation of large collection of document images. The first web-based annotation tool, WebGT \cite{biller2013webgt}, provides several semi-automatic tools for annotating degraded documents and has gained prominence recently.  An interesting prototype called Text Encoder and Annotator (TEA) is proposed in \cite{valsecchi2016text} that annotates manuscripts using semantic web technologies. 

\begin{figure}[!t]
\centering
\includegraphics[width=3.1in]{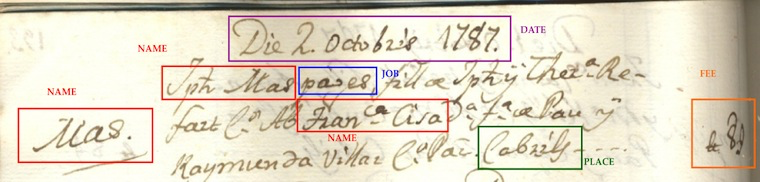}
\caption{Detailed view of the Esposalles Database \cite{romero2013esposalles}. Figure best viewed in color.}
\label{fig_barcelona_eg}
\end{figure}
 
However, these tools require specific system requirements for configuration and installation. Most of these tools and methods are either not suitable for annotating historical handwritten datasets, or represent ground truth with imprecise and inaccurate bounding boxes  \cite{wei2017use}. For example, Fig. \ref{fig_barcelona_eg} shows a detailed view of the Esposalles Database \cite{romero2013esposalles}. It can be clearly observed that the annotation for word $pages$ (blue bounding box) is inaccurate and misses certain parts of characters $p$ and $g$. Similarly, the annotation for the word $Mas$ is imprecise and the database is missing annotations for several words. This paper takes into account such issues, and proposes a simple and fast method for dynamically annotating degraded historical handwritten text on-the-fly.

\section{On-the-fly text annotation}
\label{sec:method}

\begin{figure}[!t]
\centering
\includegraphics[width=1.2in]{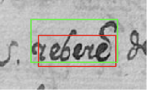}
\caption{An example illustrating the annotation for the word {\it{reber\'e}}. The red bounding box represents the user marked label. The algorithm finds the extent of the word and automatically adjusts the bounding box (green) such that the word is perfectly encapsulated. Figure best viewed in color.}
\label{fig_rebere}
\end{figure}

\begin{figure*}[!t]
\centering
\includegraphics[width=4.4in]{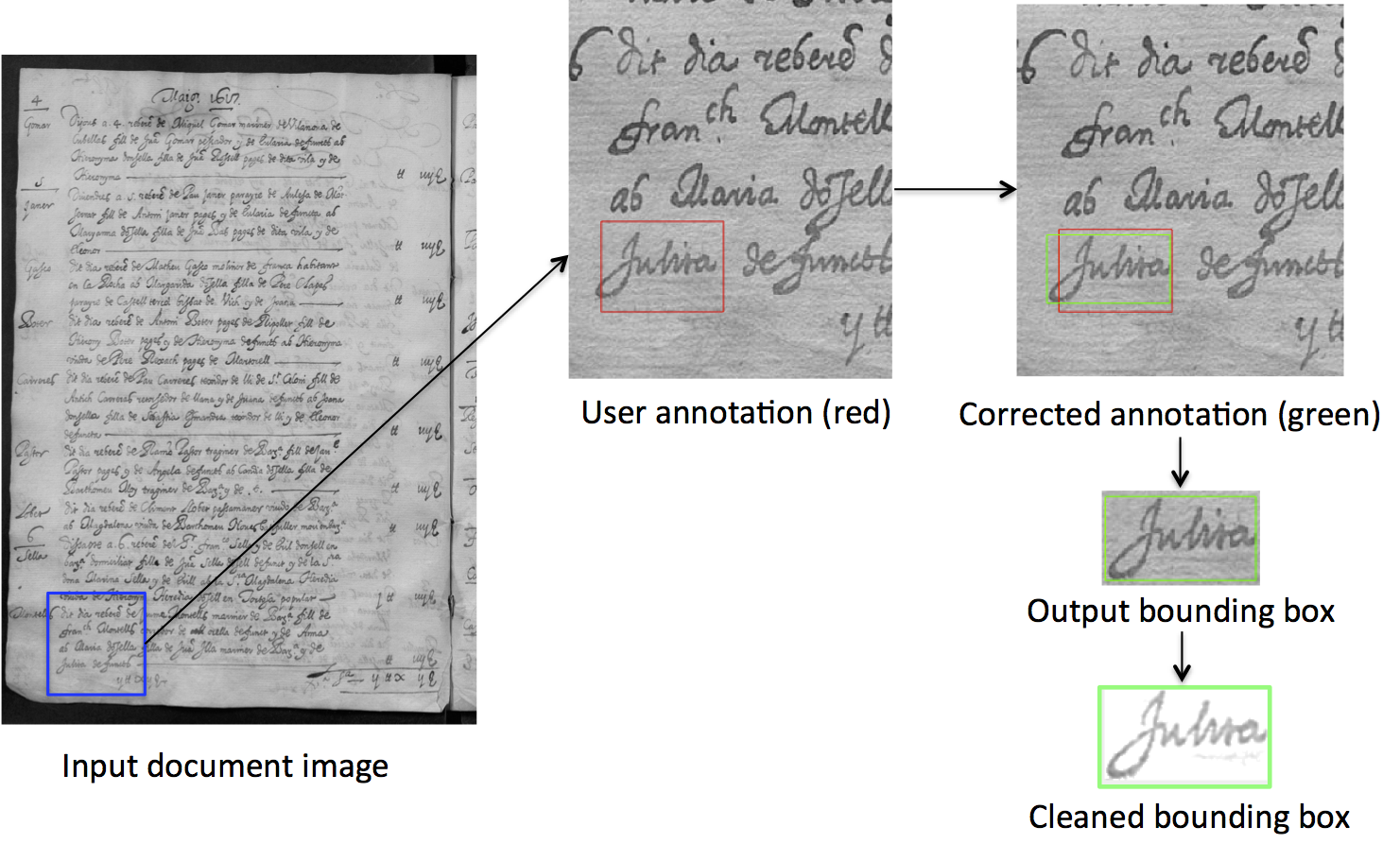}
\caption{General framework of the proposed on-the-fly text annotation method. For an input document, the user labels the query word using a simple drag-and-drop gesture. The user annotated red bounding box is then automatically adjusted and corrected. The output green bounding box represents an accurate noise-free annotation for the query word. Figure best viewed in color.}
\label{fig_meth}
\end{figure*}

The proposed on-the-fly text annotation method allows the user to generate word annotations dynamically with simple drag-and-drop gesture. The algorithm finds the extent of the word and automatically adjusts the bounding box such that the word is perfectly encapsulated and is noise-free. For example, Fig. \ref{fig_rebere} shows the user marked word in the red bounding box, and the green bounding box depicts the corrected annotation generated by the proposed method. The general framework of the proposed method is described in Fig. \ref{fig_meth}.

\begin{figure}[!t]
\centering
\includegraphics[width=1.2in]{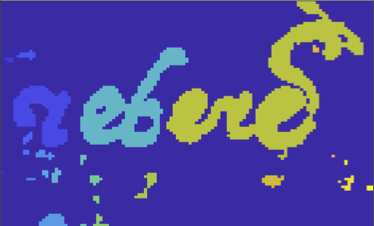}
\caption{Example of connected components extraction and labeling of the word. Figure best viewed in color.}
\label{fig_rebere2}
\end{figure}

\begin{figure}[!t]
\centering
\includegraphics[width=1.3in]{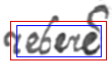}
\caption{Example of a case with blue bounding box selection that lies $\sim$5\% inside the example word. The red bounding box denotes the user annotated label. Figure best viewed in color.}
\label{fig_rebere3}
\end{figure}

The algorithm begins with the extraction of a larger bounding box area than the user selected region in order to be able to capture strokes that go beyond the selected borders. The height of the selected word is chosen as a base for area calculation. This is because the height of the selected word is approximately same across all the words in a line. The width of the selected word is also taken into account as the number of characters may vary in different words. The algorithm enhances the bounding box by one-third increase in height along the x-axis and two-third increase in height along the y-axis. 

In order to separate the word from noisy background, Gaussian filtering based background removal is performed. This is followed by connected components extraction from the word image, as highlighted in Fig. \ref{fig_rebere2} for an example word {\it{reber\'e}}. There is also some noise around the word that needs to be removed in order to avoid error in the calculation of bounding box for the extent of the word. This is done by determining whether each individual component covers a large enough area inside the word. Fig. \ref{fig_rebere3} depicts a case where the blue bounding box is selected that lies $\sim$5\% inside the example word. If a connected component covers more than 1\% of the total area inside the rectangle region, it is considered to be a part of the word. This is particularly important for the text with strokes that extend beyond the marked rectangle. Strokes from the adjacent lines and words should not be included in the bounding box even if they are partially inside the labeled region.

One can observe that the rightmost component {\it{(er\'e)}} in Fig. \ref{fig_rebere2} contains both foreground strokes and some bleed through artifacts. The noise due to bleed through is efficiently removed by thresholding the final result before generating the corrected bounding box annotation. 

\section{Experiments}
\label{sec:experiments}

This section describes the datasets used in the experiments and empirically evaluates the results obtained from the proposed word annotation algorithm.

\subsection{Dataset}

The experiments were performed on the following two publicly available datasets of varying complexity. 

\subsubsection{Esposalles dataset from the BH2M Database}
A subset of the Barcelona Historical Handwritten Marriages (BH2M) database \cite{fernandez2014bh2m} i.e. the Esposalles dataset \cite{romero2013esposalles} is used for experiments. BH2M consists of 244 books with information on 550,000 marriages registered between 15th and 19th century. The Esposalles dataset consists of historical handwritten marriages records stored in archives of Barcelona cathedral, written between 1617 and 1619 by a single writer in old Catalan. 

In total, there are 174 pages handwritten by a single author corresponding to volume 69, out of which 50 pages are selected from 17th century. The proposed method generates a document page query and allows the user to label the desired word with simple drag-and-drop gesture. For each word annotated by the user, a bounding box corresponding to the corrected annotation is generated as ground truth dynamically. 

\subsubsection{KWS-2015 Bentham dataset}
The KWS-2015 Bentham dataset, prepared in the $tranScriptorium$ project \cite{sanchez2013transcriptorium}, is employed to test the effectiveness of the proposed method. It is a challenging dataset consisting of 70 handwritten document pages and 15,419 segmented word images from the Bentham collection. The documents have been written by different authors in varying styles, font-sizes, with crossed-out words. The Bentham collection contains historical manuscripts on law and moral philosophy handwritten by Jeremy Bentham (1748-1832) over a period of 60 years, and some handwritten documents from his secretarial staff. This dataset was used in the ICDAR2015 competition on keyword spotting for handwritten documents (KWS-2015 competition) \cite{puigcerver2015icdar2015}. 

\subsection{Experimental Results}

\begin{table*}[!t]
 \centering 
 \caption{Experimental results obtained on page 2 from the Esposalles dataset after performing bounding box correction for 10 sample query words. Relative correction (\%) denotes the percentage of difference in area between the user annotated bounding box and the corrected bounding box, relative to the larger bounding box. Area and correction values are in pixels, GT denotes the ground truth.}
 \resizebox{18cm}{!} { 
  \begin{tabular}{|c|c|c|c|c|c|c|}
	\hline 
	Query word & Original GT area (O) & User annotated area (U) & Corrected GT area (C) & Correction = $\mid$O-C$\mid$ & Correction = $\mid$U-C$\mid$ & Relative correction (\%)\\
	\hline 
	reber\'e & 6608 & 7749 & 5350 & 1258 & 2399 & 30.95\\
	fill & 4956 & 4400 & 4200 & 756 & 200 & 4.54\\
	Anlefa & N/A & 8208 & 7410 & - & 798 & 9.72\\
	pages & 3960 & 6596 & 5278 & 1318 & 1318 & 19.98\\
	franca & N/A & 6765 & 8384 & - & 1619 & 19.31\\
	Candia & N/A & 4284 & 3496 & - & 788 & 18.39\\
	habitant & 8568 & 10336 & 8058 & 510 & 2278 & 22.03\\
	popular & N/A & 6681 & 9520 & - & 2839 & 29.82\\
    parayre & N/A & 6681 & 7080 & - & 399 & 5.63\\
    eleonor & N/A & 7040 & 5520 & - & 1520 & 21.59\\
	\hline
  \end{tabular}
  }
 \label{tab:barcelona1}
\end{table*}

\begin{table*}[!t]
 \centering 
 \caption{Experimental results obtained after performing bounding box correction on 10 sample query words from the KWS-2015 Bentham dataset. Area and correction values are in pixels.}
 \resizebox{18cm}{!} { 
  \begin{tabular}{|c|c|c|c|c|c|c|c|}
	\hline 
	Page \# & Query word & Original GT area (O) & User annotated area (U) & Corrected GT area (C) & Correction = $\mid$O-C$\mid$ & Correction = $\mid$U-C$\mid$ & Relative correction (\%)\\
	\hline 
	 2 & cannot & 9384 & 14490 & 8832 & 552 & 5658 & 39.04\\
	 4 & subject & 17556 & 25419 & 17264 & 292 & 8155 & 32.08\\
	 6 & whereas & 15096 & 23310 & 13400 & 1696 & 9910 & 42.51\\
	 6 & power & N/A & 6027 & 9936 & - & 3909 & 39.34\\
	 7 & number & 15768 & 31185 & 14355 & 1413 & 16830 & 53.96\\
	 10 & Law & N/A & 9944 & 15903 & - & 5964 & 37.50\\
	 13 & though & 17301 & 27776 & 17301 & 0 & 10475 & 37.71\\
	 13 & knowlege & N/A & 14040 & 27405 & - & 13365 & 48.76\\
     17 & between & 13980 & 23716 & 12095 & 1885 & 11621 & 49.00\\
     20 & demand & N/A & 11880 & 11834 & - & 46 & 0.38\\
	\hline
  \end{tabular}
  }
 \label{tab:bentham1}
\end{table*}


In order to evaluate the performance of the proposed word annotation method, bounding box labeling correction is calculated as the difference in area (in pixels) between the original ground truth available from dataset, and the method corrected bounding box. Let the areas corresponding to the original bounding box and the corrected bounding box be $O$ and $C$, respectively. The labeling correction is defined as:
\begin{equation}
\label{eq:error1}
Correction = \mid O - C \mid
\end{equation}

In the ideal case, the correction is 0. A higher value corresponding to the correction indicates inaccuracy and imprecision in the corresponding ground truth bounding box.

Since the proposed method allows users to generate missing ground truth annotations on-the-fly, correction in user generated labels is also taken into account. Therefore, the correction can also be defined as the percentage of difference in area (in pixels) between the user annotated bounding box and the corrected bounding box, relative to the larger bounding box, as follows:
\begin{equation}
\label{eq:error2a}
Correction = \mid U - C \mid
\end{equation}
\begin{equation}
\label{eq:error2b}
\textit{Relative Correction (\%)} = \Bigg\vert \frac{U - C}{max(U,C)} \Bigg\vert \times 100
\end{equation}
where, $U$ is the area corresponding to the user annotated bounding box. The percentage of relative correction quantitatively signifies the word annotation correction achieved by the proposed method. 

Table \ref{tab:barcelona1} and Table \ref{tab:bentham1} present word labeling correction results corresponding to original ground truth from the dataset and user annotated ground truth. In Table \ref{tab:barcelona1}, it can be observed that ground truth is unavailable in the Esposalles dataset for the query words $Anlefa$, $franca$, $Candia$, $popular$, $parayre$ and $eleonor$ from page 2. However, ground truth is dynamically generated for these words by user annotation on-the-fly with automatic correction using the proposed method. Similarly, Table \ref{tab:bentham1} presents the efficacy of the proposed method with reference to the KWS-2015 Bentham dataset where ground truth is unavailable for the query words $power$ on page 6, $Law$ on page 10, $knowlege$ on page 13 and $demand$ on page 20. Table \ref{tab:barcelona2} and Table \ref{tab:bentham2} present rectangle bounding box coordinates, [x-coordinate y-coordinate width height], for ground truth from dataset, user annotated ground truth, and the proposed method corrected ground truth for test query words. 

\begin{table}[!t]
 \centering 
 \caption{Experimental results representing the ground truth bounding box generated on random page 2 from the Esposalles dataset with 4 test words. BB refers to Bounding Box, denoted as [x-coordinate y-coordinate width height] of the rectangle.}
 \resizebox{9cm}{!} { 
  \begin{tabular}{|c|c|c|c|}
	\hline 
	Word & Original BB & User annotated BB & Corrected BB\\
	\hline 
	reber\'e & [340 1455 118 56] & [339 1463 123 63] & [348 1459 107 50] \\
	fill & [1097 2082 84 59] & [1050 1458 80 55] & [1057 1464 70 60] \\
	pages & [803 981 88 45] & [805 980 97 68] & [807 988 91 58]  \\
	habitant & [1103 758 168 51] & [1109 747 152 68] & [1112 757 158 51] \\
	\hline 
  \end{tabular}
  }
 \label{tab:barcelona2}
\end{table}

\begin{table}[!t]
 \centering 
 \caption{Experimental results representing the ground truth bounding box (BB) generated for 6 test words from the KWS-2015 Bentham dataset.}
 \resizebox{9cm}{!} { 
  \begin{tabular}{|c|c|c|c|}
	\hline 
	Word & Original BB & User annotated BB & Corrected BB\\
	\hline 
	cannot & [377 1027 204 46] & [369 1018 210 69] & [376 1028 192 46] \\
	subject & [306 3693 209 84] & [289 3677 229 111] & [305 3694 208 83] \\
	whereas & [542 2617 222 68] & [546 2604 222 105] & [562 2618 200 67]  \\
	number & [1147 2136 292 54] & [1144 2134 297 105] & [1157 2135 261 55] \\
    though & [201 46 219 79] & [202 22 224 124] & [202 47 219 79] \\
    between & [881 2743 233 60] & [876 2729 242 98] & [892 2745 205 59] \\
	\hline 
  \end{tabular}
  }
 \label{tab:bentham2}
\end{table}

\begin{figure}[!t]
\centering
\includegraphics[width=3.2in]{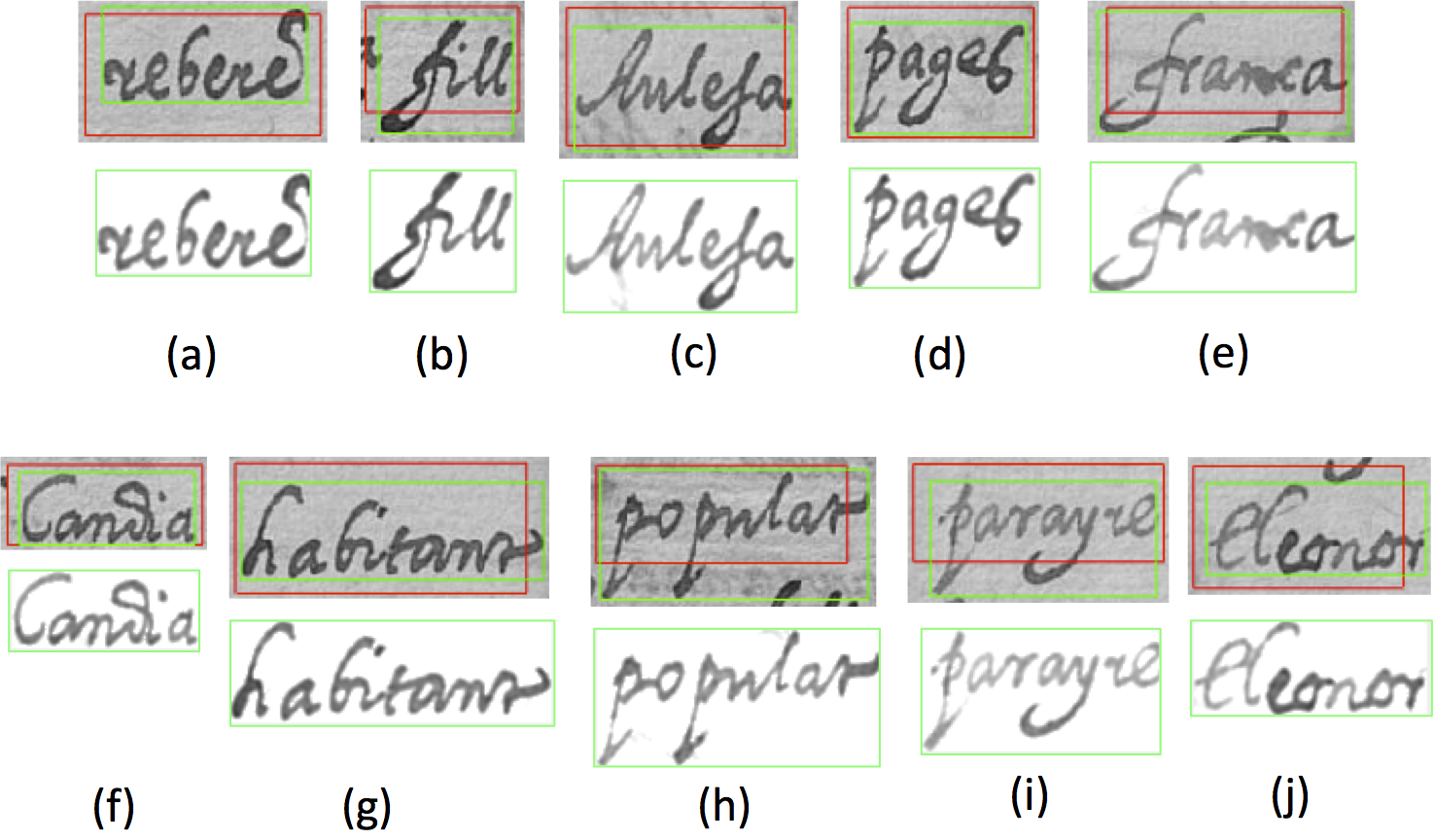}
\caption{Word annotation results on the Esposalles dataset with 10 sample query words. Red bounding box corresponds to user annotated labels. Green bounding box corresponds to method generated corrected labels. The final output is a cleaned word with accurate annotation. Figure best viewed in color.}
\label{fig_res1}
\end{figure}

\begin{figure}[!t]
\centering
\includegraphics[width=3.2in]{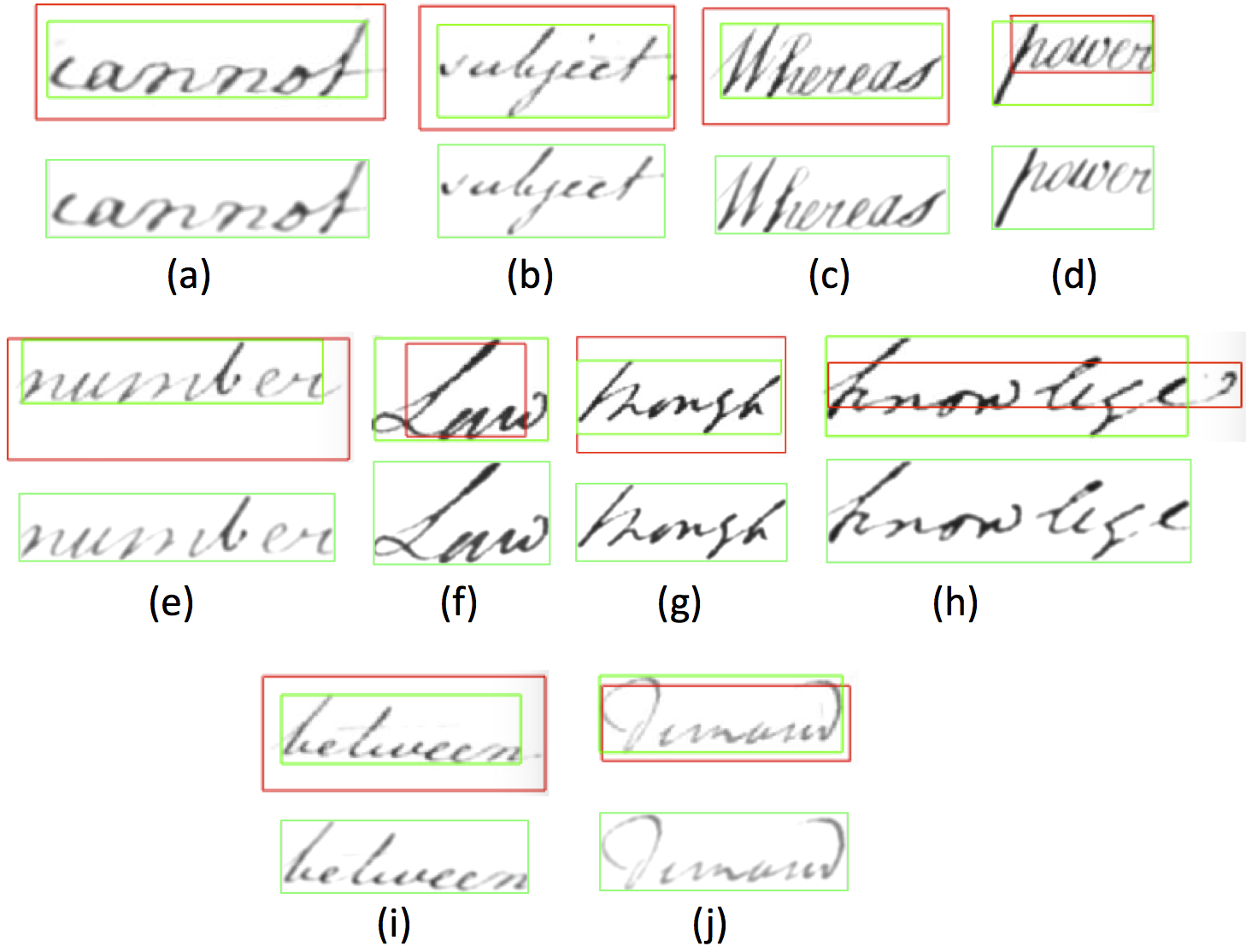}
\caption{Word annotation results on preprocessed document images from the KWS-2015 Bentham dataset with 10 sample query words. Red bounding box corresponds to user annotated labels. Green bounding box corresponds to method generated corrected labels. Figure best viewed in color.}
\label{fig_res2}
\end{figure}

Word annotation results on both datasets are visualised in Fig. \ref{fig_res1} and Fig. \ref{fig_res2}. The impact of the proposed word annotation method is demonstrated by the difference in area covered by the user annotated bounding box (in red) and the method corrected bounding box (in green). The proposed method performs background removal on word images in order to capture a noise free accurate representation of the query word. 

The proposed method performs efficiently for the document images from the Esposalles dataset, as can be seen in Fig. \ref{fig_res1}. However, the algorithm will fail if the user generated bounding box is too large, and captures multiple words and lines. 

There are some test failure cases to be taken into account with reference to the KWS-2015 Bentham dataset. To a certain extent, the proposed method struggles in identifying very fine and thin pen strokes on a degraded paper. This is because thresholding is applied on the final result before generating the corrected bounding box annotation to remove the noise due to bleed through. The thresholding also removes very thin segments of the word by mistaking it as noise. Therefore, the proposed algorithm sometimes misses certain minor details of the word. Some of the test failure cases are highlighted in Fig. \ref{fig_res2}, using the preprocessed images for better visualisation of the challenging cases. For example, in Fig. \ref{fig_res2}(h), the proposed method missed a small segment of the character $e$ in the word $knowlege$. The authors believe that this issue can be taken into account by using automatic thresholding for noise removal. 

\section{Conclusion}
\label{sec:conclusion}

A simple and efficient method to perform on-the-fly annotation of degraded historical handwritten text is presented in this paper. The novelty lies in dynamic generation of noise-free ground truth bounding box labels and automatic correction of inaccurate annotations for text in degraded historical documents. The experimental results on well-known datasets containing high levels of degradation demonstrate the effectiveness of the proposed method. As future work, automatic thresholding and parameter optimisation will be added to the algorithm, and the performance will be tested for the generation of character annotations. The ideas presented herein will be scaled to aid generation and correction of ground truth annotations for heavily degraded archival databases.

\section*{Acknowledgment}
This work has been partially supported by the eSSENCE strategic collaboration on eScience and the Riksbankens Jubileumsfond (Dnr NHS14-2068:1).



\bibliographystyle{IEEEtran}
\bibliography{refs}

\begin{thebibliography}{10}
\providecommand{\url}[1]{#1}
\csname url@samestyle\endcsname
\providecommand{\newblock}{\relax}
\providecommand{\bibinfo}[2]{#2}
\providecommand{\BIBentrySTDinterwordspacing}{\spaceskip=0pt\relax}
\providecommand{\BIBentryALTinterwordstretchfactor}{4}
\providecommand{\BIBentryALTinterwordspacing}{\spaceskip=\fontdimen2\font plus
\BIBentryALTinterwordstretchfactor\fontdimen3\font minus
  \fontdimen4\font\relax}
\providecommand{\BIBforeignlanguage}[2]{{%
\expandafter\ifx\csname l@#1\endcsname\relax
\typeout{** WARNING: IEEEtran.bst: No hyphenation pattern has been}%
\typeout{** loaded for the language `#1'. Using the pattern for}%
\typeout{** the default language instead.}%
\else
\language=\csname l@#1\endcsname
\fi
#2}}
\providecommand{\BIBdecl}{\relax}
\BIBdecl

\bibitem{pletschacher2010page}
S.~Pletschacher and A.~Antonacopoulos, ``The page (page analysis and
  ground-truth elements) format framework,'' in \emph{Pattern Recognition
  (ICPR), 2010 20th International Conference on}.\hskip 1em plus 0.5em minus
  0.4em\relax IEEE, 2010, pp. 257--260.

\bibitem{heroux2007automatic}
P.~H{\'e}roux, E.~Barbu, S.~Adam, and {\'E}.~Trupin, ``Automatic ground-truth
  generation for document image analysis and understanding,'' in \emph{Document
  Analysis and Recognition, 2007. ICDAR 2007. Ninth International Conference
  on}, vol.~1.\hskip 1em plus 0.5em minus 0.4em\relax IEEE, 2007, pp. 476--480.

\bibitem{antonacopoulos2006ground}
A.~Antonacopoulos, D.~Karatzas, and D.~Bridson, ``Ground truth for layout
  analysis performance evaluation,'' in \emph{International Workshop on
  Document Analysis Systems}.\hskip 1em plus 0.5em minus 0.4em\relax Springer,
  2006, pp. 302--311.

\bibitem{kanungoa2001trueviz}
T.~Kanungoa, C.~H. Leea, J.~Czorapinskib, and I.~Bellab, ``Trueviz: a
  groiindtruth/metadata editing and visualizing toolkit for ocr,'' in
  \emph{Proceedings of SPIE}, vol. 4307, 2001, p.~2.

\bibitem{yacoub2005perfectdoc}
S.~Yacoub, V.~Saxena, and S.~N. Sami, ``Perfectdoc: A ground truthing
  environment for complex documents,'' in \emph{Document Analysis and
  Recognition, 2005. Proceedings. Eighth International Conference on}.\hskip
  1em plus 0.5em minus 0.4em\relax IEEE, 2005, pp. 452--456.

\bibitem{saund2009pixlabeler}
E.~Saund, J.~Lin, and P.~Sarkar, ``Pixlabeler: User interface for pixel-level
  labeling of elements in document images,'' in \emph{Document Analysis and
  Recognition, 2009. ICDAR'09. 10th International Conference on}.\hskip 1em
  plus 0.5em minus 0.4em\relax IEEE, 2009, pp. 646--650.

\bibitem{doermann2010gedi}
D.~Doermann, E.~Zotkina, and H.~Li, ``Gedi-a groundtruthing environment for
  document images,'' in \emph{Ninth IAPR International Workshop on Document
  Analysis Systems (DAS 2010)}, 2010.

\bibitem{clausner2011aletheia}
C.~Clausner, S.~Pletschacher, and A.~Antonacopoulos, ``Aletheia-an advanced
  document layout and text ground-truthing system for production
  environments,'' in \emph{Document Analysis and Recognition (ICDAR), 2011
  International Conference on}.\hskip 1em plus 0.5em minus 0.4em\relax IEEE,
  2011, pp. 48--52.

\bibitem{biller2013webgt}
O.~Biller, A.~Asi, K.~Kedem, J.~El-Sana, and I.~Dinstein, ``Webgt: An
  interactive web-based system for historical document ground truth
  generation,'' in \emph{Document Analysis and Recognition (ICDAR), 2013 12th
  International Conference on}.\hskip 1em plus 0.5em minus 0.4em\relax IEEE,
  2013, pp. 305--308.

\bibitem{valsecchi2016text}
F.~Valsecchi, M.~Abrate, C.~Bacciu, S.~Piccini, and A.~Marchetti, ``Text
  encoder and annotator: an all-in-one editor for transcribing and annotating
  manuscripts with rdf,'' in \emph{International Semantic Web
  Conference}.\hskip 1em plus 0.5em minus 0.4em\relax Springer, 2016, pp.
  399--407.

\bibitem{wei2017use}
H.~Wei, M.~Seuret, M.~Liwicki, and R.~Ingold, ``The use of gabor features for
  semi-automatically generated polyon-based ground truth of historical document
  images,'' \emph{Digital Scholarship in the Humanities}, p. fqx012, 2017.

\bibitem{giotis2017survey}
A.~P. Giotis, G.~Sfikas, B.~Gatos, and C.~Nikou, ``A survey of document image
  word spotting techniques,'' \emph{Pattern Recognition}, vol.~68, pp.
  310--332, 2017.

\bibitem{yanikoglu1998pink}
B.~A. Yanikoglu and L.~Vincent, ``Pink panther: a complete environment for
  ground-truthing and benchmarking document page segmentation,'' \emph{Pattern
  Recognition}, vol.~31, no.~9, pp. 1191--1204, 1998.

\bibitem{romero2013esposalles}
V.~Romero, A.~Forn{\'e}S, N.~Serrano, J.~A. S{\'a}Nchez, A.~H. Toselli,
  V.~Frinken, E.~Vidal, and J.~Llad{\'o}S, ``The esposalles database: An
  ancient marriage license corpus for off-line handwriting recognition,''
  \emph{Pattern Recognition}, vol.~46, no.~6, pp. 1658--1669, 2013.

\bibitem{fernandez2014bh2m}
D.~Fern{\'a}ndez-Mota, J.~Almaz{\'a}n, N.~Cirera, A.~Forn{\'e}s, and
  J.~Llad{\'o}s, ``Bh2m: The barcelona historical, handwritten marriages
  database,'' in \emph{Pattern Recognition (ICPR), 2014 22nd International
  Conference on}.\hskip 1em plus 0.5em minus 0.4em\relax IEEE, 2014, pp.
  256--261.

\bibitem{sanchez2013transcriptorium}
J.~A. S{\'a}nchez, G.~M{\"u}hlberger, B.~Gatos, P.~Schofield, K.~Depuydt, R.~M.
  Davis, E.~Vidal, and J.~De~Does, ``transcriptorium: a european project on
  handwritten text recognition,'' in \emph{Proceedings of the 2013 ACM
  symposium on Document engineering}.\hskip 1em plus 0.5em minus 0.4em\relax
  ACM, 2013, pp. 227--228.

\bibitem{puigcerver2015icdar2015}
J.~Puigcerver, A.~H. Toselli, and E.~Vidal, ``Icdar2015 competition on keyword
  spotting for handwritten documents,'' in \emph{Document Analysis and
  Recognition (ICDAR), 2015 13th International Conference on}.\hskip 1em plus
  0.5em minus 0.4em\relax IEEE, 2015, pp. 1176--1180.

\end{thebibliography}
%

\end{document}